# Widefield Nanodiamond Quantum Sensing Based on Light-Sheet Microscopy

Shuo Wang[1,3,4,5+], Ming-Zhong Ai[1,3,4,5+], Jing-Wei Fan[1,2], Junchen Ye[1,3,4,5], Chao Lin[1], Quan Li[1,3,4*], Ren-Bao Liu[1,3,4,5*]

[+]These authors contributed equally: Shuo Wang, Ming-Zhong Ai

[*]Correspondence and requests for materials should be addressed to Q.L. (email: liquan@cuhk.edu.hk) or R.-B.L. (email: rbliu@cuhk.edu.hk)

1. Department of Physics, The Chinese University of Hong Kong, Shatin, New Territories, Hong Kong, China
2. School of Physics, Hefei University of Technology, Hefei, Anhui 230009, China
3. Centre for Quantum Coherence, The Chinese University of Hong Kong, Shatin, New Territories, Hong Kong, China
4. The State Key Laboratory of Quantum Information Technologies and Materials, The Chinese University of Hong Kong, Shatin, New Territories, Hong Kong, China
5. New Cornerstone Science Laboratory, The Chinese University of Hong Kong, Shatin, New Territories, Hong Kong, China

# ABSTRACT

Nanodiamonds containing nitrogen-vacancy (NV) centers are promising quantum sensors for biological applications thanks to their sub-micron spatial resolution, biocompatibility, and versatile multi-modal responses. However, the optically detected magnetic resonance (ODMR) measurement requires laser irradiation, creating a trade-off between high-throughput and low phototoxicity for applications in live cells. Here to address this challenge we develop a widefield quantum sensing method based on light-sheet microscopy (LSM), in which the sample is illuminated by a vertically movable laser sheet and the fluorescence is collected along the vertical axis that is orthogonal to the light sheet. This LSM-ODMR system is demonstrated to feature high throughput sensing due to the wide-field configuration, fast three-dimensional imaging and sensing due to the vertical mobility of the light sheet, enhanced sensitivity due to suppression of out-of-focus background fluorescence, and low phototoxicity for bio-sensing due to elimination of out-of-focus illumination. This LSM-based widefield nanodiamond sensing provides an approach for biological studies with low phototoxicity, offering three-dimensional and multi-modal sensing capability.

# INTRODUCTION

Nanodiamonds (NDs) containing nitrogen-vacancy (NV) centers are promising quantum sensors for nanoscale biological studies, owing to their multi-modal responses (to temperature, magnetic field, electric field and pressure) [1, 2], good spin coherence under ambient conditions [1], chemical inertness [3], high fluorescence stability [4], good biocompatibility [4, 5], sub-micron spatial resolution, and rich surface functionalization [6]. Previous studies have demonstrated the versatility of NV-based sensing in biological research [7, 8], including nano thermometry [9-13], intracellular orientation tracking [14-16], sub-cellular magnetic imaging [17-19], and physiologically relevant species detecting [20-24]. However,

NV-based sensing in live cells faces a critical trade-off between high throughput and low phototoxicity since laser excitation is required for optically detected magnetic resonance (ODMR) measurement [25], inducing inevitable phototoxicity to biological samples. The accumulated phototoxicity results in oxidative stress built-up and consequently numerous damages including mitochondria dysfunction and DNA damage before the eventual cell death [26, 27]. Confocal ODMR, with point-by-point scanning [25, 28], limits imaging speeds and data acquisition rates. Widefield ODMR offers high throughput [10, 28] and also spatial correlation of signals, but it associates with more significant phototoxicity due to larger laser dose across the entire sample volume via epi-illumination [27]. Widefield ODMR in the total-internal-reflection fluorescence (TIRF) mode can reduce sample excitation volume by confining the laser excitation to a ~100-nm evanescent layer near the surface [29, 30] and has been applied in NV-based biosensing [17-19], but it restricts sensing to superficial regions, losing 3D information and leading to complications due to the interference of surface effects with cellular functions in biological specimens.

Light-sheet microscopy (LSM) provides a solution to this trade-off through illumination with a vertically movable thin sheet of light perpendicular to the detection axis, enabling high-speed 3D live-cell imaging with low phototoxicity [31-33]. LSM achieves optical sectioning [34], eliminates out-of-focus excitation and background fluorescence, and supports volumetric scanning—advantages that surpass confocal (much higher speed), conventional widefield (no off-focus excitation and background fluorescence), and TIRF (full 3D capability) modalities. Although LSM has been widely adopted for live cell fluorescence imaging, its integration with ND-based quantum sensing in live cells has remained unexplored.

In this work, we integrate the advantages of LSM and ND quantum sensing by developing a widefield ODMR system based on LSM. We first demonstrate its capability for three-dimensional imaging and sensing with high sensitivity. For live-cell applications, the system provides two key advantages over conventional widefield ODMR. First, suppression of out-of-focus background fluorescence enhances the measurement sensitivity. Second, the elimination of off-plane excitation reduces the phototoxicity, allowing long-term three-dimensional ODMR monitoring within live cells up to one hour. In addition to ODMR monitoring, we develop a widefield $T_1$ relaxometry technique and demonstrate its use for long-term three-dimensional $T_1$ monitoring in live cells up to three hours. This LSM-based widefield ND quantum sensing platform paves the way for fast, sub-micron resolution biosensing with low phototoxicity, opening new avenues for quantitative, long-term studies of dynamic intracellular processes.

## RESULTS

**Scheme of LSM-ODMR** is shown in Fig. 1a. The excitation light is a two-dimensional light sheet within the $xy$ plane delivered from a side-mounted illumination objective (for further details of the optical path, see Supplementary Materials Fig. S1). Fluorescence is collected along the orthogonal $z$-direction through a detection objective. The fluorescence of NDs within the focal plane of the detection objective is simultaneously captured by a scientific CMOS (sCMOS) camera. The light sheet can be scanned along the $z$-direction to enable three-dimensional imaging and sensing of the sample. An antenna is used to introduce microwave to manipulate NV center spins. To avoid aberrations caused by the excitation light passing through the coverslip, adherent cell samples are positioned at a tilted angle [35]. Accordingly, the direction parallel to the coverslip is defined as $x'$, while the direction perpendicular to the coverslip is defined as $z'$.

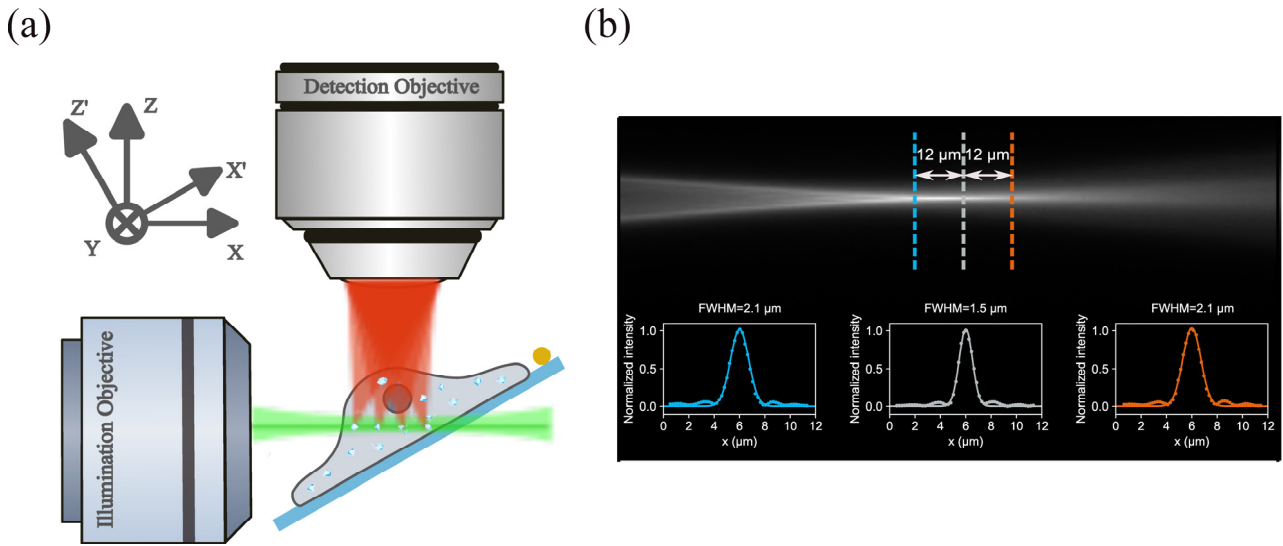

**Figure 1. Method of nanodiamond sensing based on light-sheet microscopy.** (a) Schematic of LSM-ODMR of NDs in a live cell. The green laser is introduced by the illumination objective from one side as a 2D light sheet to excite the NDs on the focal plane of the detection objective. The fluorescence of NDs is collected vertically by the detection objective. Microwave is introduced by an antenna (yellow dot) to get ODMR signal. The sample can be vertically scanned to realize 3D imaging and sensing. (b). The shape of the light sheet. The thickness is about 1.5 µm (full width at half maximum). This light sheet ensures an axial resolution of at least 2.1 µm throughout an entire HeLa cell.

Fig. 1b shows the characterization of the light sheet profile. The light sheet was visualized by exciting a 0.1% Rhodamine B dye solution and collecting its fluorescence emission. For precise characterization of the light sheet shape, a single ND sample (~100 nm in size) was used. The intensity profiles in the inset of Fig. 1b were obtained by recording the fluorescence intensities of the ND, while translating the sample (single ND attached to a coverslip) along the $z$-direction at different $x$-positions (i.e., $x = 0$, $x = -12$ µm, and $x = 12$ µm). In the non-saturated excitation regime, the fluorescence intensity of the ND is approximately proportional to the local laser intensity. Thus, the fluorescence intensity profile directly reflects the laser intensity distribution within the light sheet. The light sheet exhibited a thickness of approximately 1.5 µm, with the thickness remaining below 2.1 µm over a lateral range of 24 μm. This ensures that the axial $z$ resolution remains no worse than 2.1 μm across the dimensions of a typical HeLa cell.

The ground state of the negatively charged NV center can be optically polarized into the $m_s = 0$ spin state under green laser illumination (e.g., 532 nm). NV centers in the $m_s = 0$ state exhibit brighter fluorescence (by approximately 30%) than in the $m_s = \pm 1$ states, enabling fluorescence-based readout of the spin state. Microwave radiation resonant with spin transitions can drive the spin from the $m_s = 0$ to $m_s = \pm 1$ states, leading to a fluorescence decrease, which is the basis of optically detected magnetic resonance (ODMR) [25]. The zero-field splitting parameter $D$ between the $m_s = 0$ and $m_s = \pm 1$ levels exhibits a temperature dependence of $dD/dT \approx -74$ kHz/K [36], allowing temperature variations to be monitored through shifts of $D$. In addition to ODMR, relaxometry is a useful method for sensing magnetic noise [20], due to, e.g., spin fluctuations of paramagnetic species in the local environment, such as paramagnetic ions and reactive oxygen species (ROS) [20-24].

**3D imaging and sensing capability of LSM-ODMR system.** To demonstrate the 3D imaging and sensing performance of our system, we employed NDs embedded in agarose gel. Upon heating, the ND solution was homogeneously mixed with agarose and then cooled to form a gel. The NDs are spatially distributed throughout the agarose gel (see Methods for detailed sample preparation). The excitation laser propagates along the $x$-axis. The laser beam is collimated in $y$-direction, enabling a relatively wide field of view along this axis. The light sheet is scanned along $z$-direction. Fig. 2a displays the reconstructed 3D spatial distribution of NDs within the agarose gel. We evaluated the temperature sensitivity of NDs in the LSM-ODMR platform. As a representative result, Fig. 2b shows the standard deviation of the measured temperature against the inverse square root of the data acquisition time, and a 2.9 K/$\sqrt{\text{Hz}}$ temperature sensitivity was obtained. There is a spread in the temperature sensitivity estimated from each of the individual NDs, which is represented by the color difference in Fig. 2a. The statistical distribution of temperature sensitivity across 78 NDs (Fig. 2c) indicates that the majority exhibit sensitivity below 10 K/$\sqrt{Hz}$, with median and mean values of 4.47 K/$\sqrt{Hz}$ and 4.58 K/$\sqrt{Hz}$, respectively.

We employed ND nano-thermometry via the LSM-ODMR method to monitor the temperature changes of the agarose gel. The temperature of the dish containing the agarose-ND sample was controlled by a temperature controller and enabled by a resistance temperature detector (RTD) immediately adjacent to the sample. We systematically varied the sample temperature while detecting the change via the ODMR of the NDs. A series of temperature setpoints with a step size of approximately 3 °C were programmed. At each stabilized temperature, the light sheet was scanned across the upper five $z$-planes of the sample, with about 7-8 NDs in each of the planes and 36 NDs in total. ODMR spectra were acquired at each plane with an integration time of 100 s. The temperatures recorded by the RTD and the corresponding temperatures reported by the NDs distributed within the agarose are presented in Figure 2d. The results demonstrate the capability of system in three-dimensional imaging and simultaneous multi-point temperature measurement with high sensitivity (several K/$\sqrt{Hz}$).

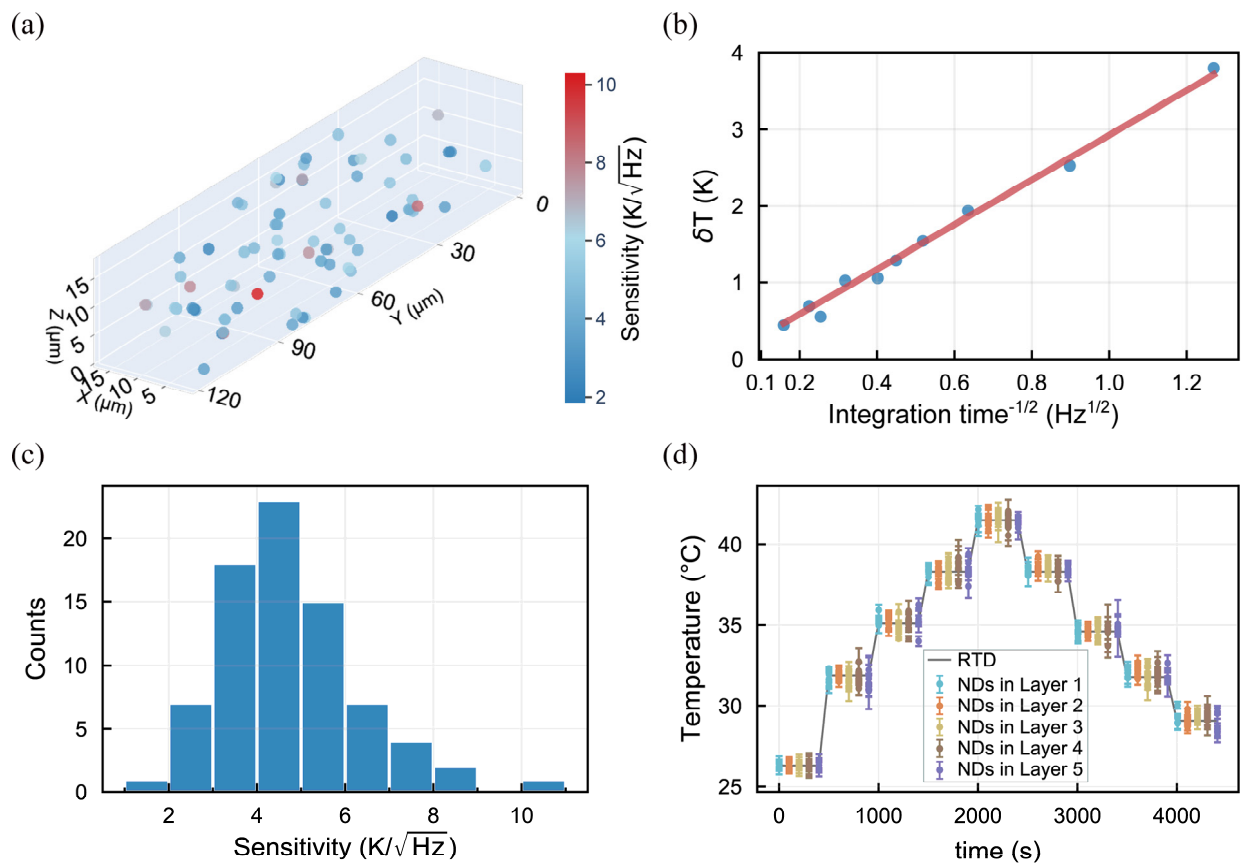

**Figure 2. 3D imaging and sensing ability of the LSM-ODMR setup.** (a) 3D imaging of NDs embedded in agarose. The laser propagates along the $x$ axis and the light sheet is scanned along the $z$ axis to realize 3D imaging. The color bar represents the range of temperature sensitivity of the NDs in the measurement volume. (b). The temperature sensitivity for a typical ND using LSM-ODMR, about 2.9 K/√Hz. (c). The temperature sensitivity distribution of the NDs in (a) using LSM-ODMR. (d). Utilizing NDs from the upper five $xy$-planes in (a) to monitor ambient temperature change. The temperature measured by an RTD adjacent to the agarose and by the NDs distributed in the agarose are shown.

**LSM-ODMR sensing in live cells.** We applied the LSM-ODMR method to live cells and demonstrated its advantages over conventional techniques. We first performed rapid 3D imaging of adherent HeLa cells and intracellular NDs using the LSM-ODMR setup (for cell sample preparation details, see Methods). The HeLa cell membrane was stained with CellMask Green dye. The 473 nm blue light sheet was used to excite the dye. The 532 nm green light sheet was applied to excite the NDs. The merged $z$- stack images and the corresponding cross-sectional views ($x'y$, $x'z'$, and $yz'$) are presented in Fig. 3a and Fig. 3b, respectively.

We validated the enhancement in measurement sensitivity achieved by the LSM-ODMR method over conventional widefield ODMR in the complex intracellular environment. For intracellular ODMR measurements, it is usually necessary to stain the whole cell or specific organelles to locate the positions of the NDs [9, 17, 23]. Together with autofluorescence of the cell, the emission of the staining dye would form background fluorescence that interferes with the ODMR signal. The optical sectioning capability of LSM eliminates the out-of-focus fluorescence background, thereby improving the signal-to-noise ratio. We first compare ODMR spectra from NDs inside unstained HeLa cells obtained using LSM and those using conventional widefield illumination at the same power density (~5 $\mu W/\mu m^2$). A typical ODMR result is shown in Fig. 3c. The slight difference in ODMR spectra may arise from the weak autofluorescence of the cell. Subsequently, we stained the HeLa cells with BCECF dye. Although the fluorescence of BCECF is primarily below 600 nm, its emission spectrum is broad, resulting in some residual fluorescence above 650 nm, which can affect the ODMR measurements. Under the conventional widefield illumination, the entire cell is illuminated, resulting in strong background fluorescence that significantly reduces the ODMR contrast (Fig. 3d) and degrades the sensitivity. In contrast, the LSM illumination excites only a thin optical section at the focal plane, minimizing dye-derived background and leading to only a slight reduction in ODMR contrast. Fig. 3e summarizes the statistical comparison of temperature sensitivity of the same NDs under LSM and the conventional widefield excitation in BCECF-stained HeLa cells, being 4.07 K/$\sqrt{Hz}$ and 8.87 K/$\sqrt{Hz}$, respectively.

We further evaluate the phototoxicity of LSM illumination compared to conventional widefield illumination. Unlike conventional microscopy, LSM illuminates only a thin optical section of the sample.

This sectioning capability substantially reduces the photodamage. We compare the viability of HeLa cells over time under LSM and widefield illumination at identical laser power densities (~5 µW/µm$^2$). (For detailed sample preparation, see Supplementary Note 3.) The results are shown in Fig. 3f. Propidium iodide (PI) dye was added to the culture medium. PI is not membrane permeable when the cells are alive. When the cells die and lose their membrane integrity, PI can enter the cells, bind to DNA, and emit strong red fluorescence, indicating cell death [37]. Cells were illuminated using either LSM or widefield, with each $xy$ plane focused continuously for approximately 50 seconds to acquire an ODMR signal with sufficient signal-to-noise ratio before moving to the next $xy$-plane for 3D measurement. Under widefield illumination, all 11 monitored HeLa cells died within 20 minutes. In contrast, under LSM illumination, all 15 HeLa cells remained viable at 20 minutes, only one cell died by 60 minutes, and all cells eventually died by 100 minutes. Representative cell images from this experiment are provided in Supplementary Materials Fig. S5. Additional comparison of cell viability using the vital dye calcein-AM under LSM versus widefield illumination is presented in Supplementary Materials Fig. S6 and Note 4. These results demonstrate that at the same laser power density, HeLa cells survived significantly longer under LSM illumination than under widefield illumination, confirming the reduced phototoxicity of LSM-ODMR.

The low phototoxicity of LSM-ODMR enables prolonged sensing in live cells. We demonstrated long-term 3D temperature monitoring in live cells using LSM-ODMR. During the experiment, the temperature of the sample dish was adjusted via the temperature controller while the ODMR signals from the intracellular NDs were continuously monitored. Each plane was continuously illuminated for 50 seconds to acquire a full ODMR spectrum with sufficient signal-to-noise ratio (the ODMR fitting error for most NDs was within 1 K), before shifting to the next $xy$-plane to achieve 3D sensing. A total of six $xy$ planes were scanned, containing 20 NDs. In Fig. 3g, the red solid line represents the temperature recorded by the RTD near the cell sample, while the blue solid line and shaded region represent, respectively, the average temperature derived from the ODMR spectra of intracellular NDs within each 150-second interval and the corresponding standard deviation among different NDs. Data from all six $xy$ planes were used in the analysis: within each interval, the mean and standard deviation were calculated from NDs located in three of the six planes (approximately 10 NDs), with the set of three planes alternating sequentially over time. The temperature trends obtained from the ODMR spectra of NDs closely follow that measured by the RTD. The variations among individual NDs possibly originated from the influence of the complex intracellular environment on the ND surfaces, which subsequently affected the $D$ shift [13], and variations of $dD/dT$ values among NDs. Cell viability after one hour of ODMR monitoring was confirmed using calcein-AM staining, as shown in Supplementary Materials Fig. S7.

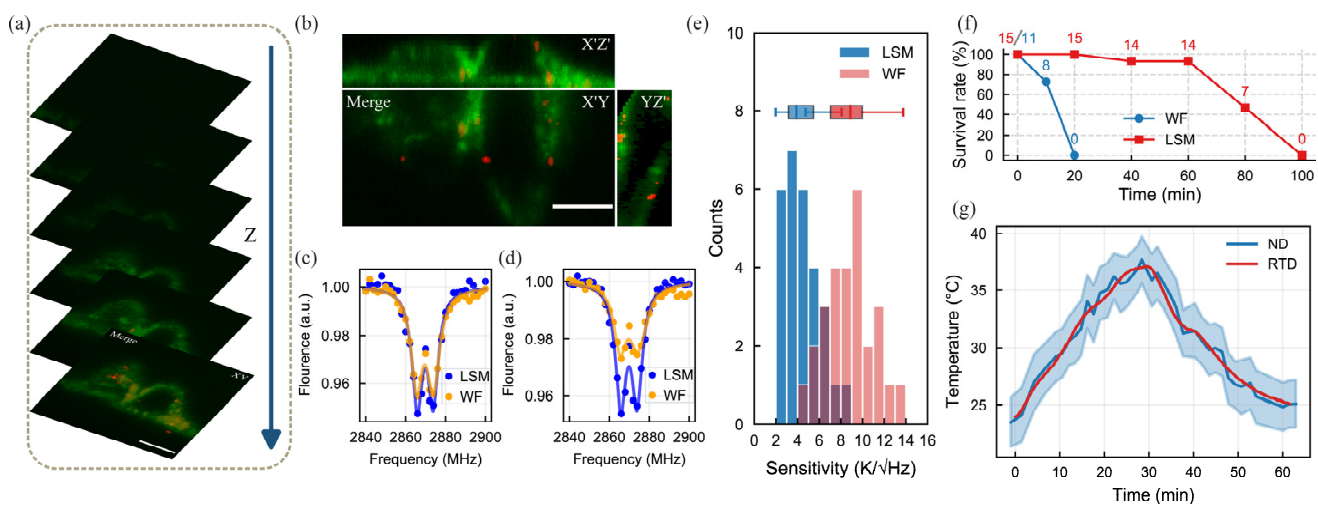

**Figure 3. Widefield nanodiamond thermometry in live cells via LSM-ODMR.** (a)-(b). $z$-stack images and $x'y$, $x'z'$ and $yz'$ cross-sectional images of intracellular NDs and HeLa cells labeled with CellMask Green dye by LSM-ODMR setup. The $x'yz'$ coordinate axes are defined in Fig. 1a. The scalebar in (a) and (b) is 10 μm. (c)-(e). Comparation of ODMR sensitivity of intracellular NDs under LSM and conventional widefield (WF) illumination. The laser power density is identical for the two illumination methods, about 5 μW/μm². (c) ODMR spectra obtained by the two illumination methods in the absence of cellular dye. (d) ODMR spectra with the addition of BCECF dye. (e) Statistical distribution of ODMR sensitivity of intracellular NDs under LSM and WF illumination (HeLa cells stained by BCECF). (f). Survival rates of HeLa cells over time under LSM and WF illumination with the same laser power density (about 5 μW/μm²). (g). Temperature monitoring of live cells using LSM-ODMR. The red line represents the temperature measured by an RTD near the cell sample. The blue curve and the blue shaded area represent the mean temperature and standard deviation, respectively, detected by NDs. Cell viability after (g) was demonstrated by vital dye calcein-AM in Supplementary Materials Fig. S7.

**LSM-$T_1$ relaxometry in live cells.** Furthermore, we developed a widefield $T_1$ relaxometry method based on LSM and applied it to live cells. Fig. 4a illustrates the pulse sequences used for LSM-$T_1$ measurement. The short laser pulse used to polarize and read out the NV center spins is set to have a duration of 5 μs, followed by a dark time $\tau_n$ ($n = 1, 2, 3, ...$). To measure the relaxation from the $m_s = \pm 1$ state, a π microwave pulse is applied immediately after the laser pulse to prepare the spin in the $m_s$

= ±1 state before relaxation. To measure the $T_1$ curve of NDs (typically several hundred microseconds to milliseconds), the dark time $\tau_n$ must be varied from the microsecond to millisecond range. As the fluorescence detector is an sCMOS camera whose shutter speed cannot reach the microsecond range, and to ensure sufficient signal per frame, each dark time $\tau_n$ is repeated $N_{\text{rep}}$ times (here, $N_{\text{rep}} = 1000$) while the camera shutter remains open. Consequently, the camera exposure time per frame ranges from milliseconds to seconds, depending on $\tau_n$. To exclude the influence of charge state conversion of NV centers on fluorescence intensity, which would otherwise interfere with $T_1$ measurements (e.g., when $NV^-$ converts to $NV^0$, the fluorescence intensity above 650 nm decreases, potentially leading to artifacts of shortened $T_1$ values), the $T_1$ signal is extracted from the difference in fluorescence intensity between the $m_s = 0$ and $m_s = \pm 1$ states. A representative $T_1$ decay curve of an ND acquired using this LSM-$T_1$ method is shown in Fig. 4b. The corresponding $T_1$ value, fitted with an exponential decay function, is 141.7±14.3 µs. For LSM-$T_1$ relaxation measurements, the $T_1$ times of NDs within a single two-dimensional plane can be acquired simultaneously. Three-dimensional mapping is then achieved by scanning the light sheet along the $z$-direction.

  We then performed long-term continuous monitoring of $T_1$ of NV center spins in NDs inside live cells using the LSM-$T_1$ relaxometry. Each $xy$ plane was continuously measured for about 270 s to obtain a reliable $T_1$ signal before refocusing the light sheet to the next $xy$ plane. Compared to the continuous-wave laser illumination used in ODMR measurements, the pulsed laser sequences employed in $T_1$ relaxometry induce lower photodamage, thereby enabling longer monitoring within live cells. Continuous $T_1$ monitoring was carried out in live HeLa cells for up to three hours. Fig. 4c shows the $T_1$ values of individual NDs tracked over this period. The $T_1$ acquisition for five $xy$ planes for the live cell constituted one measurement cycle (15 NDs in total). Every ND within the three-dimensional distribution was measured over eight such cycles. In Fig. 4d, the box plots summarize the distribution of all ND $T_1$ values for each round. Throughout the entire three-hour monitoring period, no significant change in $T_1$ values were detected. Cell viability after the measurement was confirmed by calcein-AM staining, which yielded fluorescence only in live cells (see Supplementary Materials Fig. S7). LSM-$T_1$ relaxometry enables three-dimensional monitoring of the $T_1$ times of intracellular NDs with minimal sample perturbation, thereby offering a potential approach for long-term, spatially resolved tracking of radical levels within live cells.

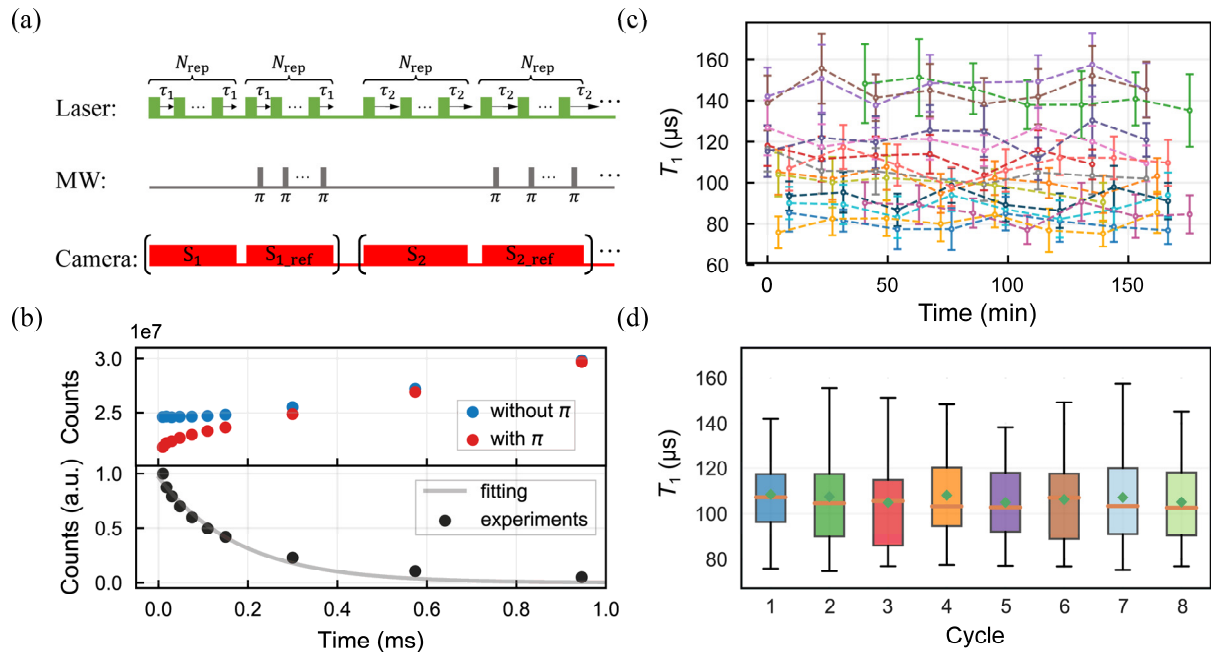

**Figure 4. Widefield nanodiamond relaxometry in live cells based on LSM.** (a). Pulse sequences used for LSM-$T_1$ relaxation measurement. The $T_1$ result is extracted by the difference of the fluorescence of $m_s = 0$ and $m_s = \pm 1$. (b). A typical $T_1$ result using LSM-$T_1$ method. The top subplot presents the raw data for relaxation from $m_s = 0$ and $m_s = \pm 1$ states. The bottom subplot shows the resulting $T_1$ curve. The $T_1$ value is 141.7±14.3 μs fitted with an exponential decay curve. (c)-(d). Monitoring $T_1$ values of NDs in a live HeLa cell using LSM-$T_1$ method. (c) shows the $T_1$ value of each ND monitored over a long period. (d) shows the distribution of $T_1$ across all NDs. Over a period of approximately three hours, eight LSM-$T_1$ measurements were performed for each ND in each two-dimensional plane within the cell. Cell viability after (c)-(d) was demonstrated by vital dye calcein-AM in Supplementary Materials Fig. S7.

## CONCLUSIONS

We have developed a widefield nanodiamond quantum sensing system based on light-sheet microscopy and demonstrated its capabilities for three-dimensional imaging and sensing. Compared to conventional methods, the LSM-based approach significantly enhances measurement sensitivity within live cells by effectively suppressing background fluorescence and thus enables more precise and reliable quantification of intracellular parameters. Moreover, the phototoxicity is drastically reduced by eliminating out-of-focus excitation. Both features contribute to reducing the laser-induced perturbation to live cells during sensing and extending their viable full-volume monitoring window to hours. Such prolonged viability potentially enables the investigation of extended physiological processes, such as tracking temperature or radical concentration dynamics at subcellular resolutions throughout the ~1-hour mitotic (M) phase of HeLa cell division [38]. Notably, the three-dimensional simultaneous multi-ND measurement rules out the chance results that may arise from single ND measurements. Furthermore, it allows for the direct spatial correlation of NDs' ODMR signals with their intracellular locations, an advantage that could be further enhanced by functionalizing ND surfaces to target specific organelles.

Gaussian light sheets face a fundamental trade-off between thickness and length due to the diffraction limit. Bessel [39] and Airy [40] beams overcome this by maintaining a longer sheet, but introduce prominent side lobes that cause out-of-focus excitation and phototoxicity. More critically for nanodiamond sensing, conventional Bessel and Airy "light sheets" are scanned virtual light sheets: the beam is focused into a line and rapidly scanned to form a 2D plane, which reduces the duty cycle for excitation and fluorescence collection, thus prolonging ODMR acquisition time. A real, non-scanning Airy light sheet [41] offers a promising alternative for future LSM-ODMR implementations.

Overall, the substantially reduced phototoxicity of our light-sheet-based approach minimizes perturbation to the biological sample and enables prolonged, reliable intracellular monitoring, thereby paving the way for quantitative studies of dynamic processes within living systems.

## METHODS

### Preparation of agarose-embedded ND samples

20 μg/ml ND aqueous solution was mixed with 2% agarose solution at a 1:1 volume ratio under continuous stirring while heating to 85–90°C. The mixture was then poured into a cubic mold, and the agarose was gelled at room temperature to embed the NDs. The final mass fraction of agarose is 1%, with a refractive index of about 1.33, close matching that of water so it causes negligible aberration. The 3D distribution of the NDs was mapped by scanning the light sheet in the $z$ axis while collecting their ODMR spectra. During variable-temperature experiments, the temperature of the agarose sample was controlled by a temperature controller (Okolab) and monitored by an RTD placed in direct contact with the sample.

**Preparation of cell samples**

A 10 × 10 mm coverslip was ultrasonically cleaned in ethanol, followed by plasma cleaning. A copper wire was then attached to the coverslip using polydimethylsiloxane (PDMS) to serve as a microwave delivery antenna. Subsequently, the coverslip was sterilized and placed at the bottom of a sterile, commercially available confocal imaging dish. HeLa cells were seeded onto the coverslip at an appropriate density in Dulbecco's Modified Eagle's Medium (DMEM, Gibco), supplemented with 10% fetal bovine serum (FBS), 0.1 g/L streptomycin sulfate, and 0.06 g/L penicillin G. The cells were incubated at 37 °C with humidity and $CO_2$ level controlled to allow adherence and growth to the desired confluency. The culture medium was then replaced. NDs (brFND-100, FND Biotech) were dispersed in fresh DMEM at a concentration of 2 μg/mL and incubated with the cells for 1 hour to allow for cellular uptake via endocytosis. Then the excess NDs were washed away with phosphate-buffered saline (PBS), followed by the addition of fresh culture medium. The cells were further incubated overnight. Then the coverslip with the cells was transferred into the LSM-ODMR sample dish like Supplementary Materials Fig. S2(b) and mounted to the LSM-ODMR setup. Throughout the experiments, the sample environment (temperature, $CO_2$ level and humidity) was maintained using controllers (Okolab), as illustrated in Supplementary Materials Fig. S1.

**3D imaging of HeLa cells and intracellular NDs**

HeLa cells were stained with 1 μM CellMask Green dye (Thermofisher, C37608) for 10 minutes to label the plasma membrane. The cell membrane was imaged using the 473 nm light sheet to excite the dye, while the NDs were imaged by the 532 nm light sheet. Three-dimensional imaging was achieved by scanning the light sheet along the $z$-direction. The raw data consisted of a series of two-dimensional images ($z$-stacks) in the $xyz$ coordinate system, which were subsequently processed into a series of two-dimensional images ($z$-stacks) in the rotated $x'yz'$ coordinate system.

**Comparison of ODMR sensitivity for the same NDs under LSM and widefield illumination in BCECF-AM-stained HeLa cells**

A commonly used intracellular pH indicator and vital dye, BCECF-AM, was employed to demonstrate the impact of background fluorescence on intracellular ODMR measurements. In the experiment, the cells were initially left unstained. The ODMR sensitivity was first measured from a single intracellular ND under either LSM or widefield illumination at identical laser power density. Following this, the cells were stained with 1 μM BCECF-AM for 15 minutes. The cell-permeant BCECF-AM ester is non-fluorescent. Once inside the cell, nonspecific esterases hydrolyze this precursor, yielding the fluorescent BCECF indicator, which is membrane-impermeant and thus trapped within the cell, resulting in intracellular fluorescence. ODMR sensitivity from the same ND was then measured again under LSM and widefield illumination at the same laser power densities as before. Although the excitation peak of BCECF is around 500 nm, the 532 nm laser used for NV excitation can still excite the dye. Owing to the dye's broad emission spectrum, some fluorescence persists above 650 nm, thereby introducing background interference to the ODMR measurement.

**ACKNOWLEDGEMENTS**

This work was supported by the Research Grant Council of HKSAR Collaborative research fund (C4007-19G), Croucher foundation under project No. SRF23401, the New Cornerstone Science Foundation and the National Natural Science Foundation of China (Grant No. 92565113).

## AUTHOR CONTRIBUTIONS

R.B.L. and Q.L. conceived the idea and supervised the project, S.W., M.Z.A., J.W.F., Q.L. and R.B.L. designed the experiments, S.W., M.Z.A. and J.W.F. constructed the setup with the contribution from J.C.Y. and C.L., S.W. and M.Z.A. performed the experiments, S.W., M.Z.A, Q.L., and R.B.L. analyzed the data, S.W., M.Z.A., Q.L., R.B.L. wrote the paper, and all authors commented on the paper.

## REFFERENCES